\documentclass[preprint,12pt]{elsarticle}

\usepackage{graphicx}
\usepackage{subfigure}
\usepackage{epstopdf}
\usepackage{amsmath}
\usepackage{txfonts}
\usepackage{booktabs}

\usepackage{amssymb}
\usepackage{makecell,multirow,diagbox}

\usepackage{bm}

\usepackage{amsmath,amsfonts,amssymb,graphicx}    

\begin{document}

\begin{frontmatter}

\title{Grid-Forming Converters control based on DC voltage feedback}

\author[label{a}]{Yuan Gao\corref{cor1}}
\author[label{a}]{Hai-Peng Ren\corref{cor1}}
\author[label{a}]{Jie Li\corref{cor1}}
\address[label{a}] {Shaanxi Key Laboratory of Complex System Control and Intelligent Information Processing, Xi'an University of Technology, Xi'an, Shaanxi, 710048, China}


\begin{abstract}
The renewable energy is connected to the power grid through power electronic converters, which are lack of make the inertia of synchronous generator/machine (SM) be lost. The increasing penetration of renewable energy in power system weakens the frequency and voltage stability. The Grid-Forming Converters(GFCs) simulate the function of synchronous motor through control method in order to improve the stability of power grid by providing inertia and stability regulation mechanism. This kind of converter control methods include virtual synchronous machine, schedulable virtual oscillator control and so on. These control method mainly use AC side state feedback and do not monitor the DC side state. This paper analyzes the control strategy of GFC considering power grid stability, including Frequency Droop Control, Virtual Synchronous Machine Control and dispatchable Virtual Oscillator Control. The DC side voltage collapse problem is found when a large load disturbance occurs. The control methods of GFC considering DC side voltage feedback are proposed, which can ensure the synchronization characteristics of grid connection and solve the problem of DC side voltage collapse. The proposed method is verified by IEEE-9 bus system, which shows the effectiveness of the proposed method.
\end{abstract}


\end{frontmatter}

\section{Introduction}
\label{1}
Renewable energy power generation technology provides conditions for solving the contradiction between the increasingly exhausted traditional energy and the continuous growth of energy demand [1]. The transition from fossil fuel based thermal power generation to renewable energy generation power converters connection to the grid, which leads to the weakening of inertia and the threatening to grid voltage and frequency stability control mechanism of synchronous machines. This reduction of moment of inertia makes the power system to be a low inertia network, which brings serious stability challenges to the power grid [2-4].

In order to solve this problem, grid forming converters (GFCs) technology is considered as the cornerstone of the future power system. According to the characteristics and functions of synchronous motor, the grid connected converter must support load sharing / droop characteristics, black start, inertial response and hierarchical frequency / voltage regulation functions [5]. These functions enable SMs and GFCs to ensure the stability of the system in the interaction.

In recent years, the control strategies of GFCs include Frequency Droop Control, which imitates the frequency droop regulation mechanism existing in synchronous generators, is a widely applied grid connected converters[6]. In order to further simulate the external characteristics of SMs, a virtual synchronous generator (VSG) control strategy [7-9] is proposed to make the GFC demonstrate synchronous generator external characteristics. The virtual oscillator control (VOC) simulates the synchronization behavior of Li¨¦nard type oscillator. VOC can the power system consisting of renewable energy connection using GFC be synchronized globally, but the rated power cannot be determined. dVOC (dispatchable virtual oscillator control) [11-12] can ensure synchronization and control AC power to reach the pre-calculated working point(i.e., the rated power), thus overcoming the limitations of VOC.

The existing control methods of GFCs do not consider the DC side working state, which might lead to voltage collapse caused by DC side current over limit in case of large load disturbance, other states of power system, such as power, voltage, frequency. Reference [13] proposed an AC current saturation algorithm to improve the stability of AC side voltage. But it can not solve the problem of DC voltage collapse caused by DC current over limit. To solve this problem, the DC voltage feedback is introduced to the GFC to ensure the DC side current and voltage keep in a stable range, which solves the problem of DC voltage collapse and improves the stability of the whole system.

The rest of this paper is arranged as follows: Section 1 introduces the mathematical model of GFC and analyzes the problems in existing control methods; section 2 proposes a new control strategy of GfC based on DC voltage feedback; section 3 uses IEEE-9 bus system to verify the proposed strategy; section 4 gives the conclusion.

\section{GFC model and problem statement}
\subsection{Grid forming converter}
The structure of GFC is shown in Fig. 1, including DC power supply module, inverter module, filter module and grid connected converter control module


\begin{figure*}
  \centering
  \subfigure[]{\includegraphics[width=3in]{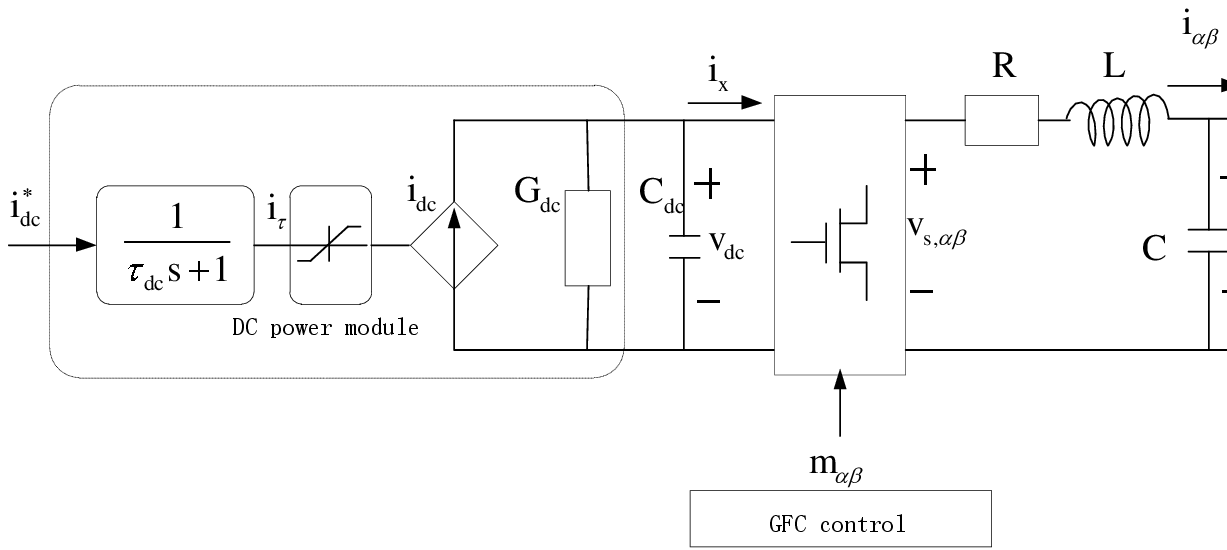}}

  \caption{ schematic structure }

\end{figure*}

The unified model of the grid connected converter is given by [14]:
\begin{eqnarray}
\begin{array}{l}
 {C_{dc}}{{\dot v}_{dc}} = {i_{dc}} - {G_{dc}}{v_{dc}} - {i_x}, \\
 L{{\dot i}_{s,\alpha \beta }} = {v_{s,\alpha \beta }} - R{i_{s,\alpha \beta }} - {v_{\alpha \beta ,}} \\
 C{{\dot v}_{\alpha \beta }} = {i_{s,\alpha \beta }} - {i_{\alpha \beta, }} \\
 \end{array}
\end{eqnarray}
Where $ {C_{dc}}$ is the DC link capacitance, ${G_ {dc}} $ is the DC side conductance, $L $, $C $, $R $ are filter inductance, capacitance and resistance, respectively; ${i_\tau }$ is the current supplied from DC source; ${i_ {dc}}$, ${v_ {dc}} $ is DC current and voltage; ${m_ {alpha / beta}} $ represents the modulation signal of the GFC; ${i_ x} = \left( {1/2} \right)m_ {\alpha \beta }^T{i_ {s, alpha}} $ is the DC current delivered to the switch; ${i_ {s, alpha / beta}} $ and ${v_ {s, alpha}} $ are the output current and voltage of the grid connected converter (before the output filter), respectively; ${i_ {alpha / beta}} $and ${v_ {alpha / beta}} $is the GFC output current and voltage, respectively.
\subsection{DC side current limitation and its problem}

In order to prevent DC side current from over-limiting caused by load disturbance, the traditional GFC adopts formula (2) to limit the DC side current

\begin{eqnarray}
{i_{dc}} = sat\left( {{i_\tau },i_{\max }^{dc}} \right) = \left\{ \begin{gathered}
  \begin{array}{*{20}{c}}
   {{i_\tau }} & {\begin{array}{*{20}{c}}
   {} & {} & {}  \\

 \end{array} \begin{array}{*{20}{c}}
   {}  \\

 \end{array} if\left| {{i_\tau }} \right| < \left| {i_{\max }^{dc}} \right|,}  \\

 \end{array}  \hfill \\
  \begin{array}{*{20}{c}}
   {sgn\left( {{i_\tau }} \right)i_{\max }^{dc}} & {if\left| {{i_\tau }} \right| \geqslant \left| {i_{\max }^{dc}} \right|,}  \\

 \end{array}  \hfill \\
\end{gathered}  \right.
\end{eqnarray}
where $i_ {max} ^ {dc} $is the maximum DC current limit amplitude.

Figure 2 shows the variation curve of DC voltage and current under virtual synchronous generator control. Figure. 2 (a) shows the change of DC side current when the load increases from 0p.u. to 0.78p.u. at 1s, and Fig. 2 (b) shows the change of DC side voltage corresponding to subgraph (a). In Fig. 2 (a), the black dotted line represents the current demand, the red solid line represents the actual DC current, and the blue dotted line represents the DC current limit. It can be seen that the stability of DC load is not affected by such load disturbance.

\begin{figure*}
  \centering
  \subfigure[]{\includegraphics[width=2in]{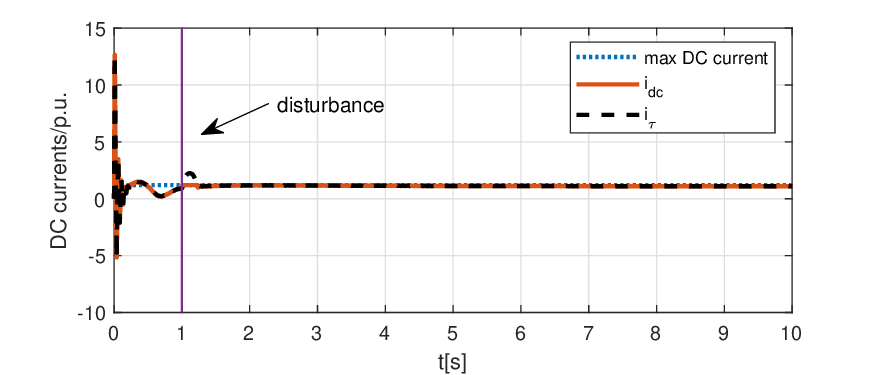}}
  \subfigure[]{\includegraphics[width=2in]{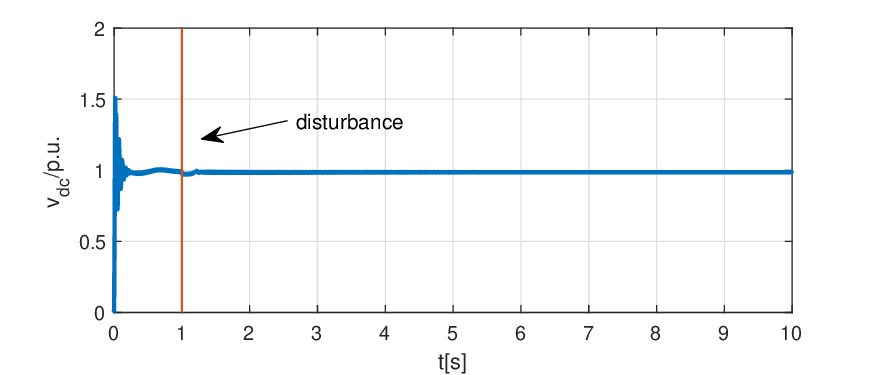}}
  \caption{ waveform of DC current (a) and voltage (b) with small load disturbance at 1s }

\end{figure*}

However, when the load disturbance increases from 0 p.u. to 0.9 p.u., Figure 3 shows the curve of DC side states under the same control strategy as that in Fig. 2.

The meanings of different line types in Fig. 3 are the same as those in Fig. 2. Figure 3 (a) shows the change of DC side current. After 4 S, the DC side current is always in saturation state, and Fig. 3 (b) shows the change of DC side voltage. It can be seen that the DC link capacitor is constantly discharging to make up for the lack current ${i_ \tau } - i_ {\max }^{dc}$.  Long time DC power saturation will lead to DC voltage collapse.
\begin{figure*}
  \centering
  \subfigure[]{\includegraphics[width=2in]{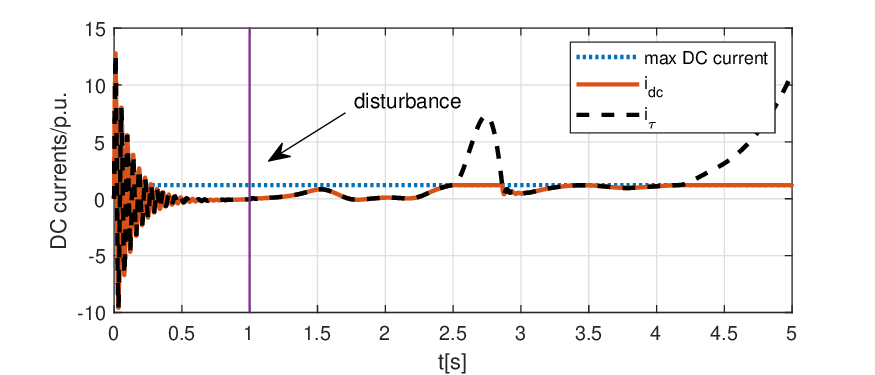}}
  \subfigure[]{\includegraphics[width=2in]{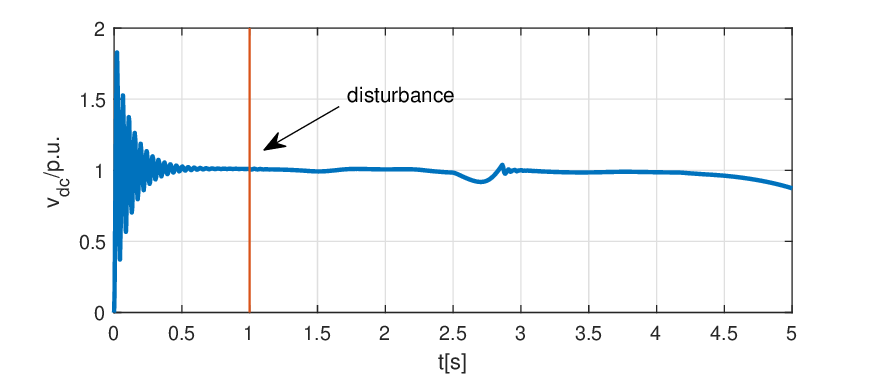}}
  \caption{waveform for large load disturbance, DC current (a), voltage (b) }

\end{figure*}

Figure 4 shows the curve of each state when the virtual synchronous generator control strategy is adopted in the corresponding AC side in Fig. 3. Figure 4 (a) shows the AC side frequency waveform and Fig. 4 (b) shows the active power waveform. It can be seen from Fig.3 and Fig.4 that when the GFC controlled by virtual synchronous generator method is disturbed with the large load disturbance, the DC side current and voltage may become unstable, which in turn leads to large fluctuations of AC side frequency, power and other states. If other GFC control methods are adopted, such as Frequency Droop Control, dVOC, the situation is similar to that in Fig. 3 and Fig. 4, which are given here.
\begin{figure*}
  \centering
  \subfigure[]{\includegraphics[width=2in]{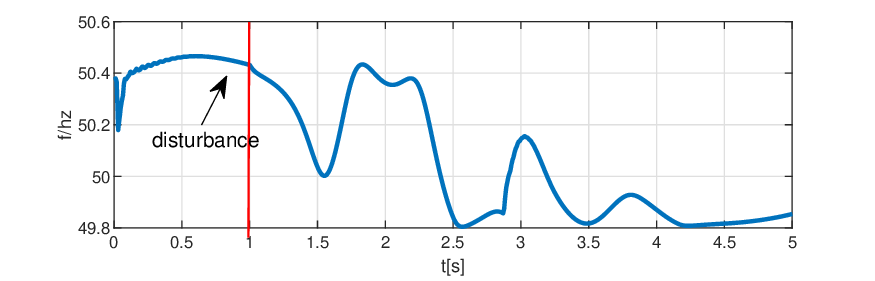}}
  \subfigure[]{\includegraphics[width=2in]{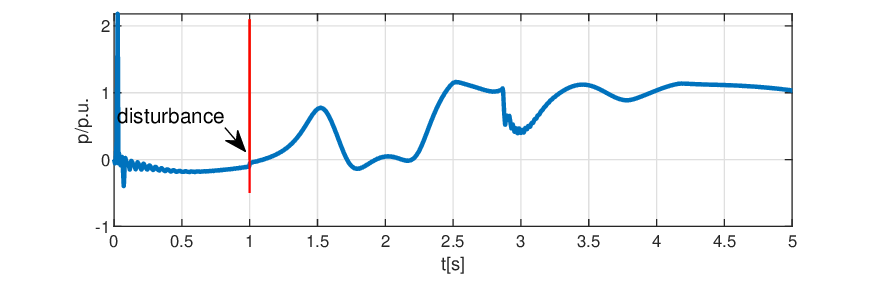}}
  \caption{waveform of AC side frequency (a), active power (b) under large load disturbance}

\end{figure*}

For the DC side voltage collapse caused by large load disturbance, the reference [13] had proposed to limit the AC current, that is, adding an AC current limiting module in to the conventional current and voltage double closed-loop control to prevent the over discharge of GFC DC link capacitor. The double closed-loop control with AC side current limitation block diagram is shown in Fig. 5

\begin{figure*}
  \centering
  \subfigure[]{\includegraphics[width=3in]{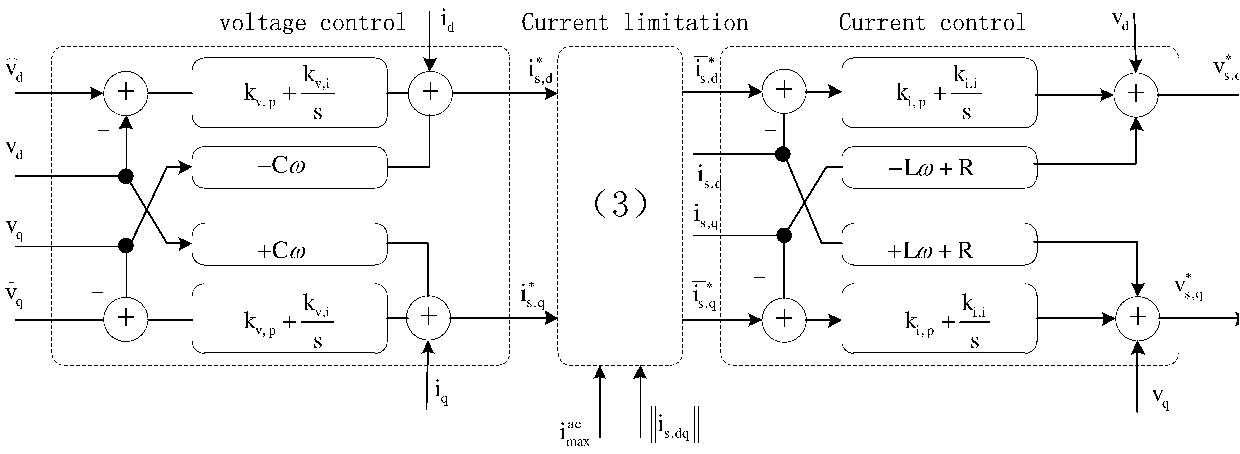}}

  \caption{ double closed loop control block diagram with AC current limiting }

\end{figure*}

In Fig. 5, AC current limitation is added on the basis of AC voltage control loop and current control loop. The AC side current limit is as follows:
\begin{eqnarray}
\bar i_{s,dq}^ *  = \left\{ \begin{gathered}
  \begin{array}{*{20}{c}}
   {i_{s,dq}^ * } & {\begin{array}{*{20}{c}}
   {}  \\

 \end{array} if\left| {{i_{s,dq}}} \right| \leqslant \left| {i_{\max }^{ac}} \right|}  \\

 \end{array}  \hfill \\
  \begin{array}{*{20}{c}}
   {{\gamma _i}i_{s,dq}^ * } & {if\left| {{i_{s,dq}}} \right| > \left| {i_{\max }^{ac}} \right|,}  \\

 \end{array}  \hfill \\
\end{gathered}  \right.
\end{eqnarray}
where $\bar i_{s,dq}^ * $ is the limited amplitude reference current holding $i_{s,dq}^ * $; ${\gamma _i} = \left( {{{i_{\max }^{ac}} \mathord{\left/
 {\vphantom {{i_{\max }^{ac}} {\left\| {i_{s,dq}^ * } \right\|}}} \right.
 \kern-\nulldelimiterspace} {\left\| {i_{s,dq}^ * } \right\|}}} \right)$ is the amplitude limitation factor. Figure 6 shows the change curve of DC side current and voltage corresponding to those in Fig. 4 when AC current limiting is added.
 \begin{figure*}
  \centering
  \subfigure[]{\includegraphics[width=2in]{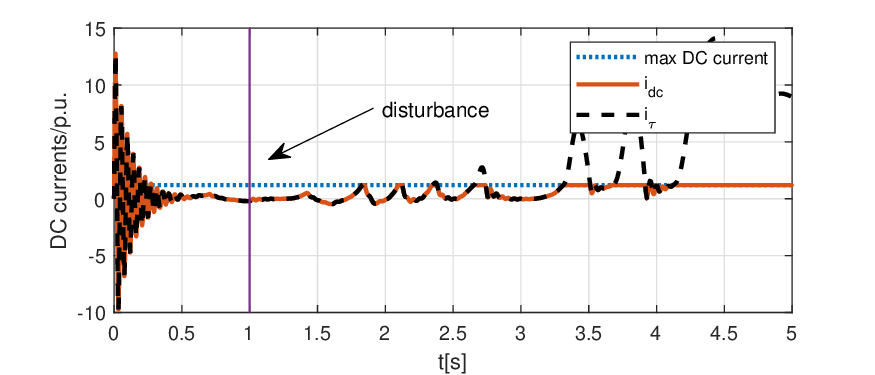}}
  \subfigure[]{\includegraphics[width=2in]{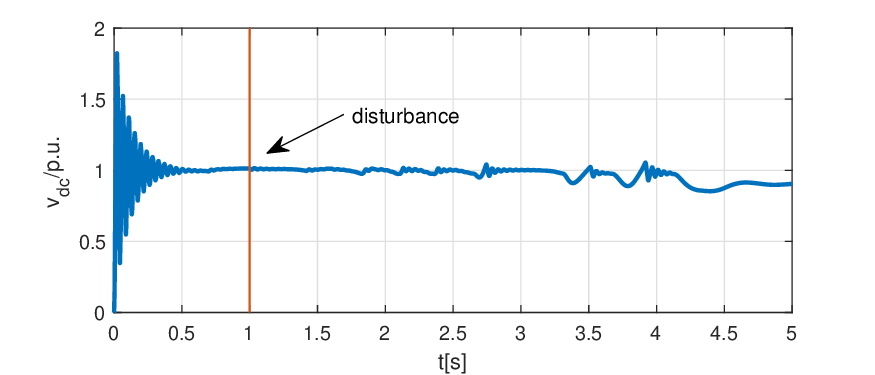}}
  \caption{DC current (a), voltage (b)}

\end{figure*}
It can be seen from Fig. 6 that the DC side voltage collapse cannot be avoided because of the long-term DC side current shortage when the VSG method with AC current limitation is added. Using other control methods such as Frequency Droop Control, dVOC could achieve the similar results as that in Fig. 6, which are not given here for limited length of the paper.

\section{GFC control considering DC voltage feedback}
 The current GFC control methods only consider the AC side performance by using the AC side state feedback control, which is the reason of the DC voltage collapse caused by the DC current shortage for long time when the large load perturbation occurs. To solve this problem, this paper proposes to add the DC voltage feedback into the traditional GFC controllers, so as to avoid the DC voltage collapse caused by large load disturbance. The DC voltage feedback is incorporated into droop control, VSG and dVOC, respectively, to show the effectiveness of the proposed method.
\subsection{VSG control with DC voltage feedback}
The VSG control is described as follows:
\begin{eqnarray}
\begin{gathered}
  \dot \theta  = \omega,  \hfill \\
  J\dot \omega  = \frac{1}
{{{\omega ^ * }}}\left( {{p^ * } - p} \right) + {D_p}\left( {{\omega ^ * } - \omega } \right), \hfill \\
\end{gathered}
\end{eqnarray}
where, ${\omega ^ * }$ is the rated frequency, $\omega $ is the actual frequency, $\theta $ is the phase angle, $p $ is the active power, ${p^ * }$ is the active power setting value, ${D_p}$ is the damping coefficient and $J $is the virtual moment of inertia. Considering adding the DC voltage feedback control term $\omega  = \frac{{{v_{dc}}}}
{{v_{dc}^ * }}{\omega ^ * }$ into the original VSG controller, Equation (4) is changed into equation (5):
\begin{eqnarray}
\begin{gathered}
  \dot \theta  = \omega,  \hfill \\
  \dot \omega  = \alpha \left( {\frac{1}
{{J{\omega ^ * }}}\left( {{p^ * } - p} \right) + \frac{{{D_p}}}{J}\left( {{\omega ^ * } - \omega } \right)} \right) \hfill \begin{array}{*{20}{c}}
     \end{array}+ \left( {1 - \alpha } \right)\frac{d}
{{dt}}\left( {\frac{{{v_{dc}}}}
{{v_{dc}^ * }}{\omega ^ * }} \right), \hfill \\
\end{gathered}
\end{eqnarray}
where ${v_{dc}}$ is the DC side voltage, $v_{dc}^ * $ is the DC side voltage setting value, and $\alpha $ is the weight to balance the traditional control and DC voltage feedback. The control block diagram of VSG with DC side voltage feedback is shown in Fig. 7. The red dotted box is the DC side voltage feedback part added in this paper:
\begin{figure*}
  \centering
  \subfigure[]{\includegraphics[width=3in]{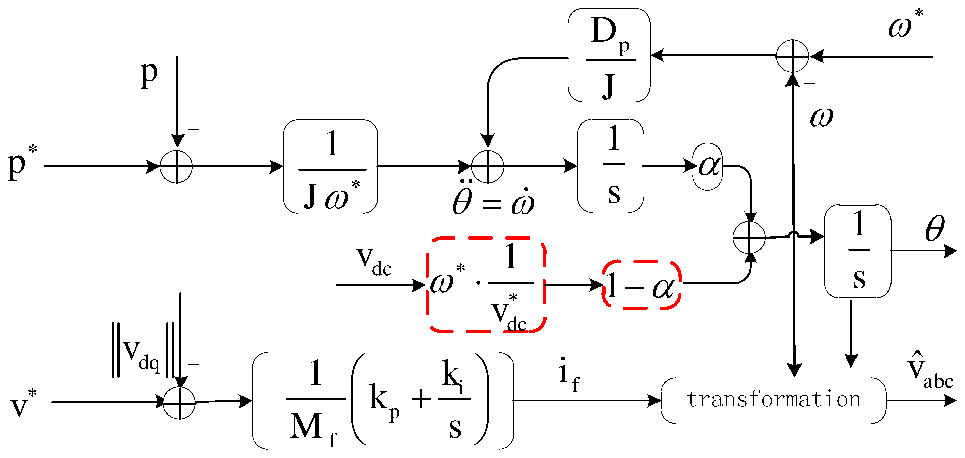}}

  \caption{ new VSG control with DC voltage feedback }

\end{figure*}

\subsection{Frequency Droop Control with DC voltage feedback}
Frequency Droop Control formula:
\begin{eqnarray}
\begin{gathered}
  \dot \theta  = \omega,  \hfill \\
  \omega  = {\omega ^ * } + {d_\omega }\left( {{p^ * } - p} \right), \hfill \\
\end{gathered}
\end{eqnarray}
where ${d_\omega }$ is the droop gain, and other variables are defined in the same as formula (4). After adding DC voltage feedback to the original Frequency Droop Control method, the following controller is obtained
\begin{eqnarray}
\begin{gathered}
  \dot \theta  = \omega,  \hfill \\
  \omega  = \alpha \left( {{\omega ^ * } + {d_\omega }\left( {{p^ * } - p} \right)} \right) + \left( {1 - \alpha } \right)\left( {\frac{{{v_{dc}}}}
{{v_{dc}^ * }}{\omega ^ * }} \right). \hfill \\
\end{gathered}
\end{eqnarray}

The control block diagram of frequency droop control with DC voltage feedback is shown in Fig. 8, where the red dotted box is the added DC voltage feedback item.
\begin{figure*}
  \centering
  \subfigure[]{\includegraphics[width=3in]{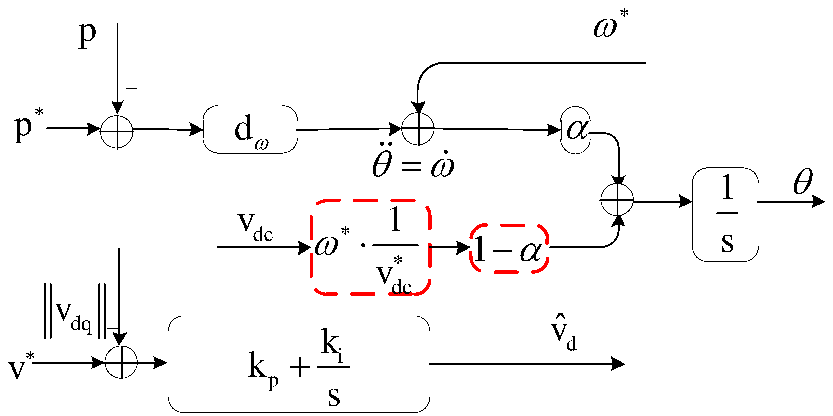}}

  \caption{ new Frequency Droop Control model with DC voltage feedback control }

\end{figure*}

\subsection{dVOC control with DC voltage feedback}
dVOC controller is as follows:
\begin{eqnarray}
\begin{gathered}
  {{\dot \theta }_i} = {\omega ^ * } + \eta \left( {\frac{{p_i^ * }}
{{v_i^{ * 2}}} - \frac{{{p_i}}}
{{{{\left\| {{v_i}} \right\|}^2}}}} \right), \hfill \\
  \left\| {{{\dot v}_i}} \right\| = \eta \left( {\frac{{q_i^ * }}
{{v_i^{ * 2}}} - \frac{{{q_i}}}
{{{v_i}}}} \right)\left\| {{v_i}} \right\| + \frac{{\eta \mu }}
{{v_i^{ * 2}}}\left( {v_i^{ * 2} - \left\| {v_i^2} \right\|} \right)\left\| {{v_i}} \right\|, \hfill \\
\end{gathered}
\end{eqnarray}
where $q $ is the reactive power, ${q^ * }$ is the set value of reactive power, $\eta$, $\mu $ are control gain, ${v_i }$ is the output reference voltage of the controller, and other variables are defined in the same formula (4). After adding DC voltage feedback to the original control method, the controlled system is given as follows:
\begin{eqnarray}
\begin{gathered}
  {{\dot \theta }_i} = {\omega ^ * } + \eta \alpha \left( {\frac{{p_i^ * }}
{{v_i^{ * 2}}} - \frac{{{p_i}}}
{{{{\left\| {{v_i}} \right\|}^2}}}} \right) + \left( {1 - \alpha } \right)\left( {\frac{{{v_{dc}}}}
{{v_{dc}^ * }}{\omega ^ * }} \right), \hfill \\
  \left\| {{{\dot v}_i}} \right\| = \eta \left( {\frac{{q_i^ * }}
{{v_i^{ * 2}}} - \frac{{{q_i}}}
{{{v_i}}}} \right)\left\| {{v_i}} \right\| + \frac{{\eta \mu }}
{{v_i^{ * 2}}}\left( {v_i^{ * 2} - \left\| {v_i^2} \right\|} \right)\left\| {{v_i}} \right\|. \hfill \\
\end{gathered}
\end{eqnarray}

The control block diagram of the dVOC with the DC voltage feedback is shown in Fig. 9, and the red dotted line box is the added DC voltage feedback item.
\begin{figure*}
  \centering
  \subfigure[]{\includegraphics[width=3in]{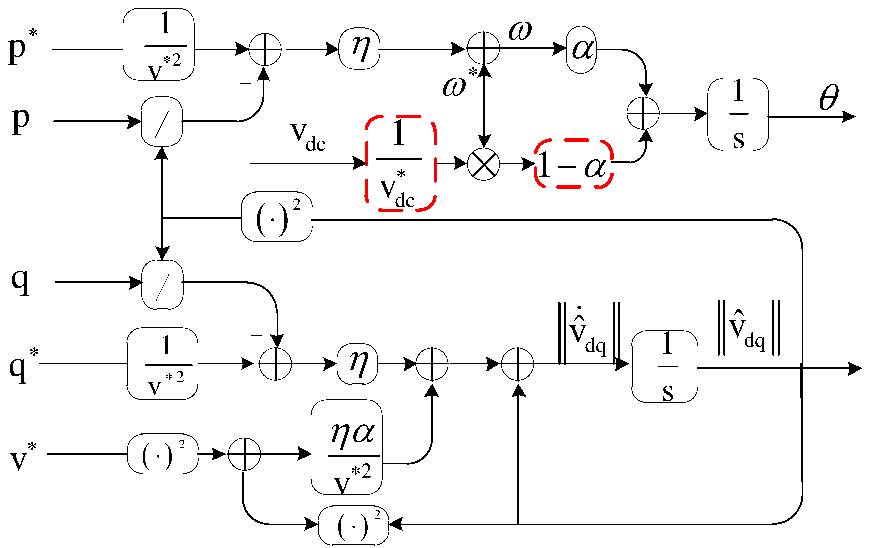}}

  \caption{ new dVOC model control with DC voltage feedback  }

\end{figure*}

\section{Simulation results and their analysis}
In this paper, IEEE-9 sections system shown in Fig. 10 is used for simulation verification. SM in the figure is the generator, GFC1 and GCF2 are two grid connected converters, which are connected to the buses through three transformers.
\begin{figure*}
  \centering
  \subfigure[]{\includegraphics[width=3in]{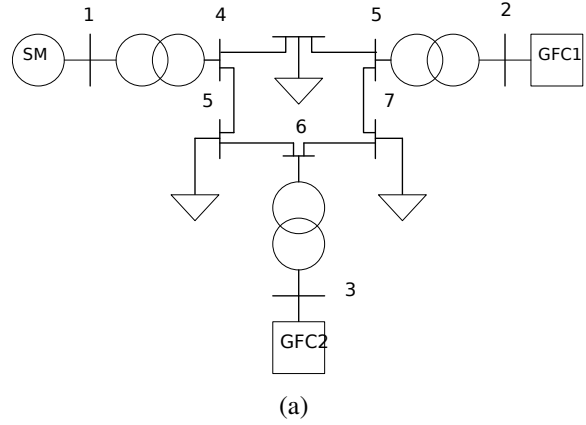}}

  \caption{ IEEE-9 sectors system model }

\end{figure*}

\subsection{Simulation parameters}
The IEEE-9 sectors test system is simulated by Matlab / Simulink. RLC model is adopted for modeling the line, and the transformer is modeled by three-phase linear transformer model. The parameters of simulation model are given in Table 1.
\begin{center}
{\fontsize{9.3pt}{11.6pt}\selectfont

\begin{tabular}{cccccc}
\hline\toprule
\multicolumn{6}{c}{ieee-9 bus test system parameters}                                                                                                                     \\ \hline\toprule
\multicolumn{1}{l}{${S_b}$} & \multicolumn{1}{l}{100MVA} & \multicolumn{1}{l}{${v_b}$} & \multicolumn{1}{l}{230kV} & \multicolumn{1}{l}{${\omega _b}$} & \multicolumn{1}{l}{$2\pi 50rad/s$} \\ \hline
\multicolumn{6}{c}{Droop control parameters}                                                                                                                             \\ \hline
${d_\omega}$    &$2\pi 0.05$   &${\omega ^ * }$     & 2$\pi$50     &${k_p},{k_i}$  & 0.001£¬0.5     \\ \hline
\multicolumn{6}{c}{VSG parameters}                                                                                                                              \\ \hline
${D_p}$         &${10^5}$     &$J$       &$2 \times {10^3}$ &${k_p},{k_i}$  &0.001£¬0.0021    \\ \hline
\multicolumn{6}{c}{dVOC parameters}                                                                                                                             \\ \hline
$\eta $           &0.021     &$\mu$     &$6.66 \times {10^4}$ &$K$  &${\pi  \mathord{\left/{\vphantom {\pi  2}}\right. \kern-\nulldelimiterspace} 2}$  \\ \hline\toprule
\end{tabular}

}
\end{center}
\subsection{Simulation parameters}

\noindent1) Frequency Droop Control

Figure 11 shows the simulation results of Frequency Droop Control method with DC voltage feedback when the load disturbance increases from 0 p.u. to 0.9 p.u. at 1s. The curves of DC side current and voltage are shown in subplots  (a) and (b), respectively. It can be seen that the DC side current can be stabilized to the rated value in a short transient time.

\begin{figure*}
  \centering
  \subfigure[]{\includegraphics[width=2in]{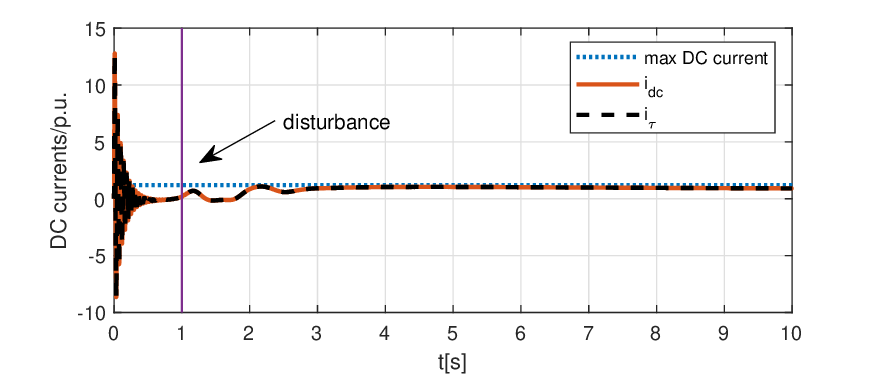}}
  \subfigure[]{\includegraphics[width=2in]{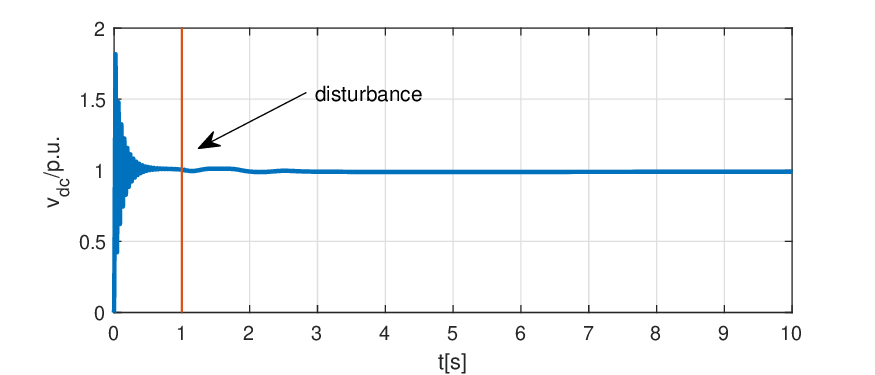}}
  \caption{ Frequency Droop Control with DC voltage feedback, DC current waveform, (b) DC voltage waveform. }
\end{figure*}
Figure 12 shows the waveforms of AC side frequency, active power and output voltage amplitude corresponding to Fig. 11. It can be seen that the Frequency Droop Control method with DC voltage feedback ensures the stability of the whole system.
\begin{figure*}
  \centering
  \subfigure[]{\includegraphics[width=2in]{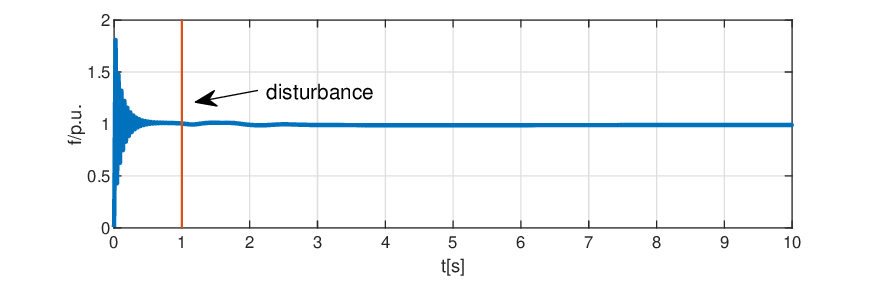}}
  \subfigure[]{\includegraphics[width=2in]{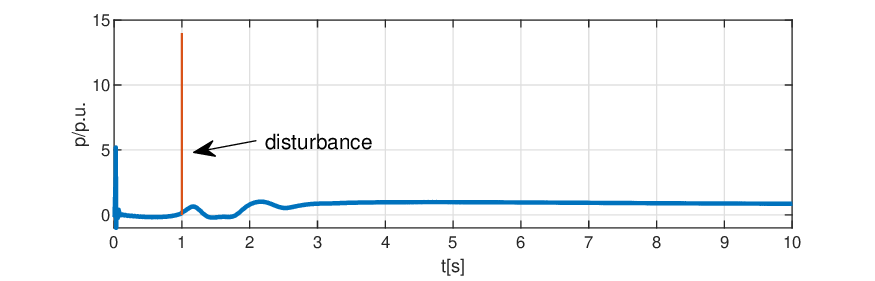}}
  \caption{ frequency (a), active power (b) corresponding to the case in Fig.11 }
\end{figure*}

\noindent2) Simulation results of VSG control

Figure. 13 shows the waveform of DC side current and DC voltage of VSG control method with DC voltage feedback when the load disturbance increases from 0 p.u. to 0.9 p.u. in steady state at 1s.
\begin{figure*}
  \centering
  \subfigure[]{\includegraphics[width=2in]{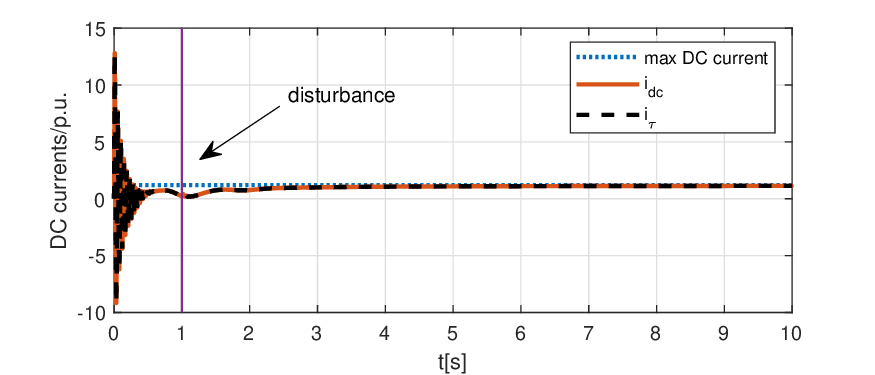}}
  \subfigure[]{\includegraphics[width=2in]{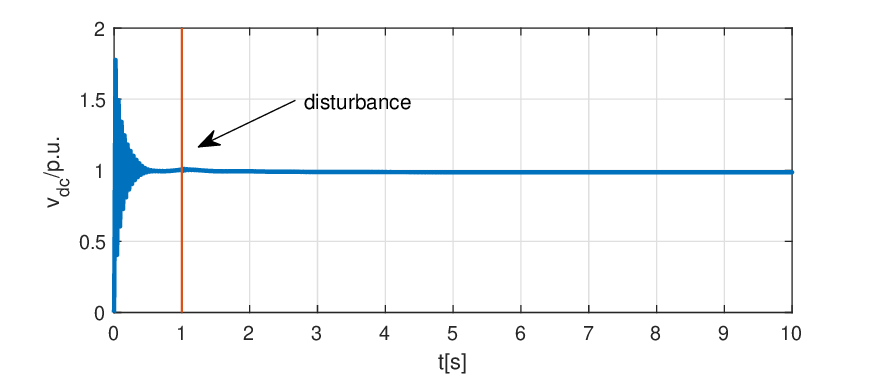}}
  \caption{ VSG control with DC voltage feedback, (a)DC current waveform, (b) DC voltage waveform }
\end{figure*}
Figure 14 shows the waveforms of AC side frequency (a), active power (b) and output voltage amplitude (c) corresponding to the situation in Fig. 13. As can be seen from Figs. 13 and 14, VSG control with DC voltage feedback ensures the stability of both DC and AC sides.
\begin{figure*}
  \centering
  \subfigure[]{\includegraphics[width=2in]{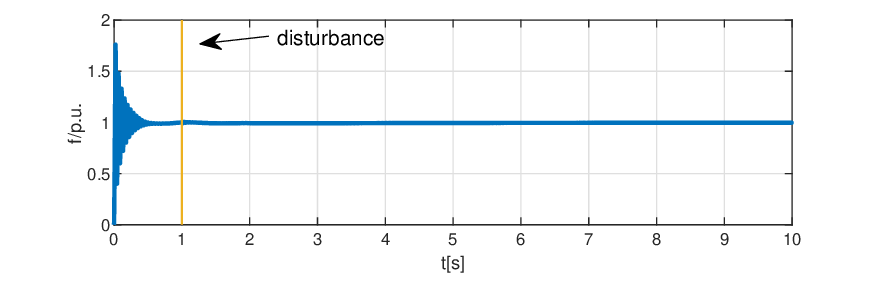}}
  \subfigure[]{\includegraphics[width=2in]{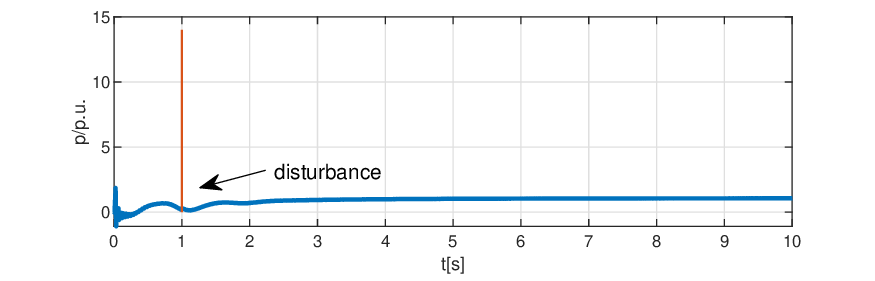}}
  \subfigure[]{\includegraphics[width=2in]{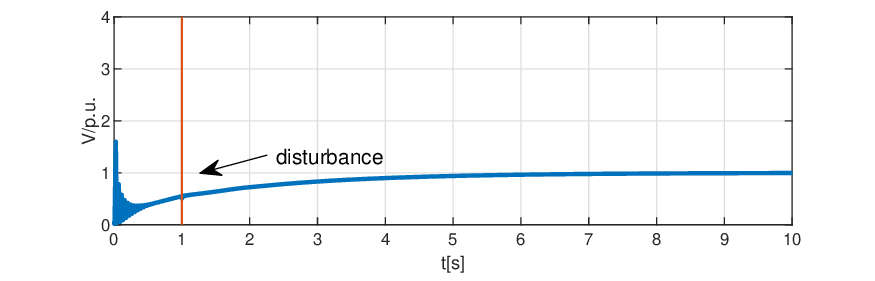}}
  \caption{ AC side frequency (a), active power (b) and output voltage amplitude (c) corresponding to the situation in Fig.13 }
\end{figure*}
\noindent3) dVOC control simulation results

 Figure 15 (a) and (b) show the waveforms of DC side current and voltage under the control of dVOC control method when the load disturbance increases from 0 p.u. to 0.9 p.u. at 1s. It can be seen from Figs. 11, 13 and 15 that by adding DC voltage feedback to the traditional grid connection control methods including FDC, VSG, and dVOC, can solve the problems of DC side current over-limited and DC voltage collapse caused by large load disturbance.
\begin{figure*}
  \centering
  \subfigure[]{\includegraphics[width=2in]{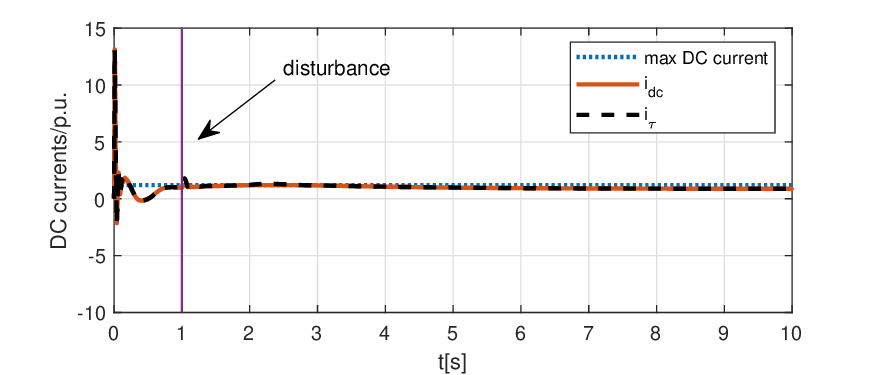}}
  \subfigure[]{\includegraphics[width=2in]{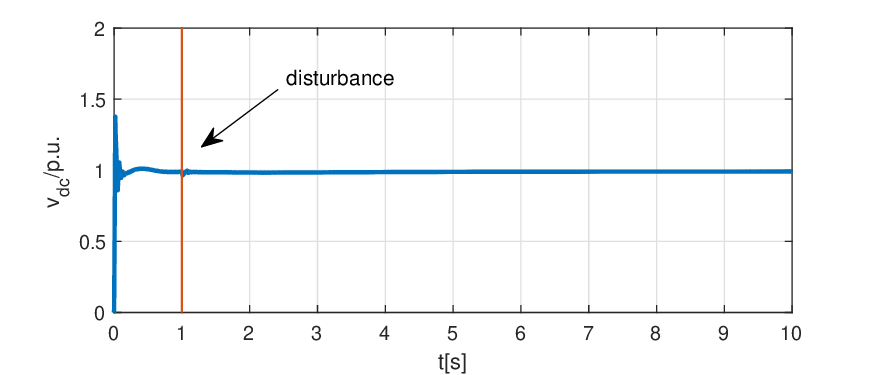}}
  \caption{ DVOC control with DC side voltage feedback, (a)DC current waveform, (b) DC voltage waveform}
\end{figure*}

Figure 16 shows the waveform of frequency (a), active power (b) and output voltage amplitude (c) corresponding to Fig. 15. It can be seen from Figs. 12, 14 and 16 that adding DC voltage feedback into the traditional grid connection control strategy can stabilize all electrical states of the system to a stable value in a short time under the condition of large load disturbance.
\begin{figure*}
  \centering
  \subfigure[]{\includegraphics[width=2in]{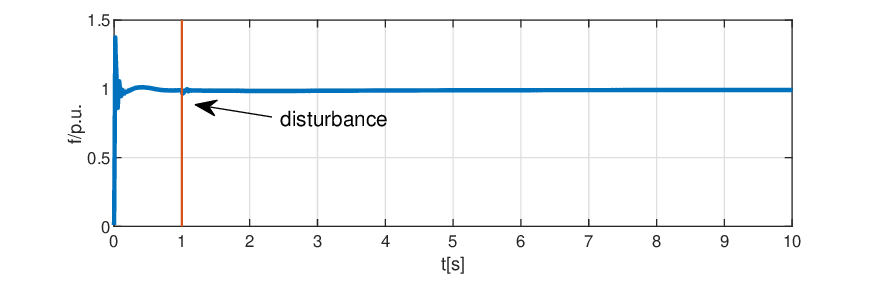}}
  \subfigure[]{\includegraphics[width=2in]{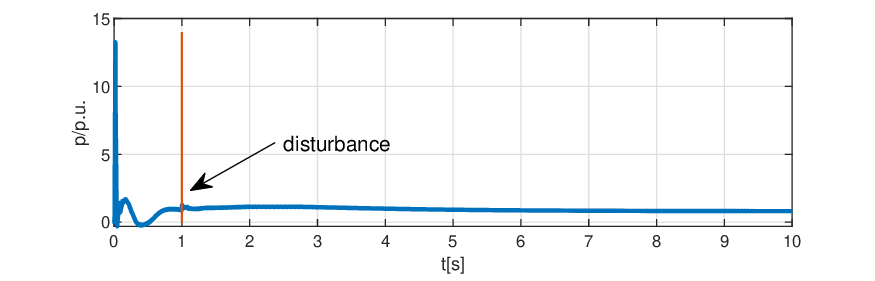}}
  \subfigure[]{\includegraphics[width=2in]{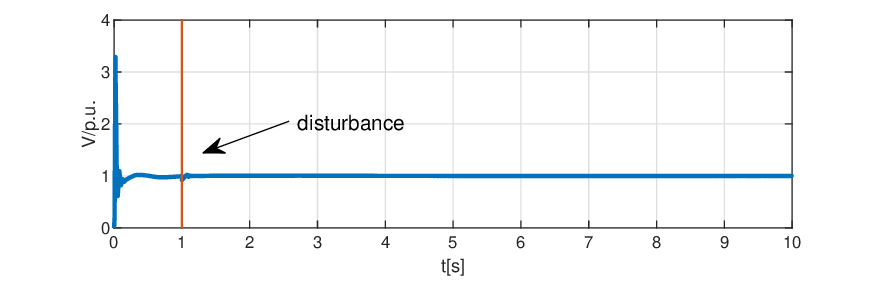}}
  \caption{ AC side frequency (a), active power (b) and output voltage amplitude (c) corresponding to the situation in Fig.15 }
\end{figure*}
Figure 17 shows the time-varying curves of frequency (a), active power (b) and output voltage amplitude (c) of SM, GFC1 and GFC2 under the VSG control method based on DC voltage feedback when the load disturbance increases from 0p.u. to 0.9p.u. it can be seen that after the system is disturbed, each electric quantity will fluctuate in varying degrees. In about 1.5s, all electrical quantities tend to be stable. The corresponding electrical quantities of other nodes are synchronized.
\begin{figure*}
  \centering
  \subfigure[]{\includegraphics[width=2in]{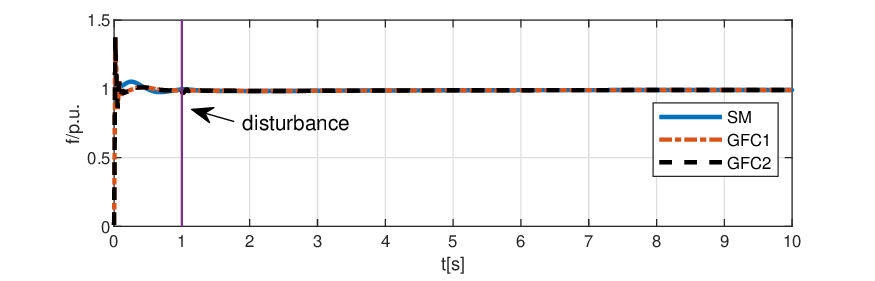}}
  \subfigure[]{\includegraphics[width=2in]{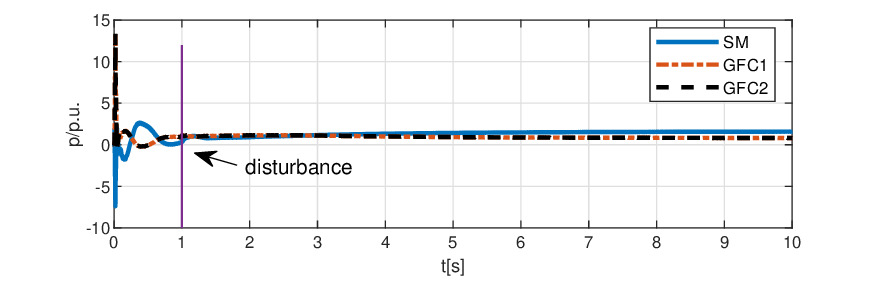}}
  \subfigure[]{\includegraphics[width=2in]{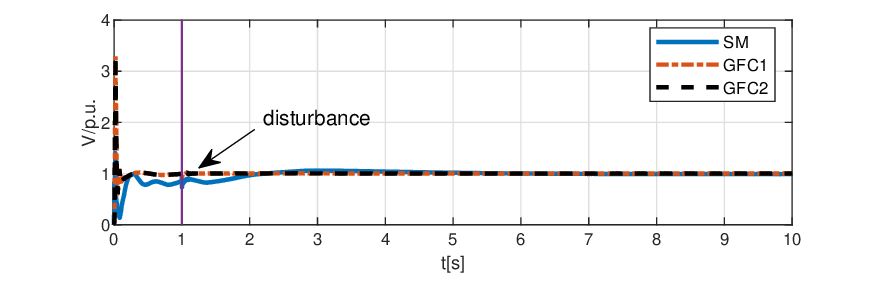}}
  \caption{ frequency (a), active power (b) and output voltage amplitude (c) on AC side }
\end{figure*}

To sum up, Frequency Droop Control, VSG and dVOC are, respectively, based on the control formula (6), (8) and (10). Since the controller only considers the measurement and feedback of AC power, although it can improve the frequency stability performance of the system, the DC side state is unknown, which might lead to DC side voltage collapse under large disturbance. On the basis of the original controller, DC voltage feedback is introduced, The formulas (7), (9) and (11) are obtained respectively, which can make the DC side and AC side be controlled coordinately, and ensure the AC side performance and the DC side stability. The simulation results show that the proposed control method improves the robustness of the system when the power system is disturbed in a large amplitude.
\section{Conclusion}

In this paper, the IEEE-9 node simulation model is built in Matlab / Simulink environment. The simulation results show that the current GFC control methods such as Frequency Droop Control, VSG and dVOC might cause DC side current overlimited and, voltage collapse, and then lead to the instability of other electrical state of the system, such as frequency, active power, etc. In this paper, the DC side voltage feedback is added to the existing GFC control methods The simulation results show that the method is effective to solve the problem stated above.

\end{document}